\documentclass[fleqn,usenatbib]{mnras}

\usepackage{newtxtext,newtxmath}

\usepackage[T1]{fontenc}

\DeclareRobustCommand{\VAN}[3]{#2}
\let\VANthebibliography\thebibliography
\def\thebibliography{\DeclareRobustCommand{\VAN}[3]{##3}\VANthebibliography}

\usepackage{graphicx}	
\usepackage{amsmath}	
\usepackage{ragged2e}

\usepackage{subcaption}

\newcommand{\kms}{km~s$^{-1}$}
\newcommand{\sbu}{mag~arcsec$^{-2}$}

\newcommand{\re}{$R_{\rm e}$}
\newcommand{\glxA}{DFUWS~68}
\newcommand{\glxB}{DFUWS~258}

\title[Quenched Field LSBGs]{The Stellar Populations of Two Quiescent Low Surface Brightness Dwarf Galaxies in Low Density Environments}

\author[D. A. Forbes et al.]{
Duncan A. Forbes$^{1}$\thanks{E-mail: duncan.forbes@gmail.com},
Hannah S. Christie$^{1}$, 
Arsen Levitskiy$^{1}$,
Jonah S. Gannon$^{2,3,1}$,
Anna Ferre-Mateu$^{4,5,1}$,
\newauthor
Aaron J.\ Romanowsky$^{6,7}$,
and Jean P. Brodie$^{1,7}$\\ 
$^{1}$Centre for Astrophysics and Supercomputing, Swinburne University of Technology, John Street, Hawthorn VIC 3122, Australia \\
$^{2}$ Department of Astronomy \& Astrophysics, University of Toronto, 50 St. George Street, Toronto, ON M5S 3H4, Canada\\
$^{3}$ Dragonfly Focused Research Organization, 150 Washington Avenue, Suite 201, Santa Fe, NM 87501, USA\\
$^{4}$ Instituto de Astrofísica de Canarias, Calle Vía Láctea S/N, E-38205 La Laguna, Tenerife, Spain\\
$^{5}$ Departamento de Astrof\'isica, Universidad de La Laguna, E-38200, La Laguna, Tenerife, Spain \\
$^{6}$Department of Physics \& Astronomy, San Jos\'e State University, One Washington Square, San Jose, CA 95192, USA\\
$^{7}$Department of Astronomy \& Astrophysics, University of California Santa Cruz, 1156 High Street, Santa Cruz, CA 95064, USA
}

\date{Accepted XXX. Received YYY; in original form ZZZ}

\pubyear{\the\year{}}

\begin{document}
\label{firstpage}
\pagerange{\pageref{firstpage}--\pageref{lastpage}}
\maketitle

\begin{abstract}
Dwarf galaxies located in low density environments that have quiescent stellar populations are extremely rare and the mechanism for quenching their stars is highly uncertain. Here we present two low surface brightness dwarf galaxies with large ($>$1.2 kpc) effective radii that were previously identified in the Dragonfly Ultrawide Survey for which we have obtained spectra using the KCWI instrument on the Keck II telescope.  
We derive their stellar populations and star formation histories in several radial bins out to 1.25 effective radii. 
Globally both galaxies are old ($\sim$9 Gyr) and metal-poor ([M/H] $\sim$ --1) and reveal quenching on a relatively rapid timescale ($\sim$ 1 Gyr). 
We find flat-to-rising age and metallicity gradients in both galaxies. This is similar to those seen in ultra diffuse galaxies but is in contrast with typical classical dwarf galaxies and predictions from simulations. 
One galaxy, \glxA, hosts several globular clusters which we find to share similar ages, metallicities and quenching timescales to that of the host galaxy stars. 
We briefly discuss possible quenching mechanisms and suggest that 
neither cosmic web stripping nor internal feedback processes alone can explain our results. 


\end{abstract}




\section{Introduction}

The difference between the theoretical halo mass function and the observed galaxy stellar mass function is due to the suppression of star formation within dark matter halos \citep{2010ApJ...710..903M}. 
This quenching of star formation is thought to be caused largely by the feedback from AGN in high mass halos and from supernova in lower mass halos. In addition to these internal processes, external processes can also play an important role. For a review of quenching mechanisms, see  
\cite{2018NatAs...2..695M}.

For dwarf galaxies, 
a variety of mechanisms have been proposed to quench them as they fall into clusters and orbit within the cluster over cosmic time, e.g. \cite{2014MNRAS.440.1934T}. 
This includes galaxies which have passed through the cluster centre and now occupy the back-splash region  in the cluster outskirts  \citep{2021NatAs...5.1255B}. However, 
the mechanisms that quench dwarf galaxies well outside of clusters, in low density environments, are less clear and poorly constrained. 
A possible reason for this is that quenched galaxies in the field are extremely rare 
\citep{2012ApJ...757...85G}.
For example, the 
recent work of 
\cite{2025ApJ...994..231K} showed that quenched fractions 
approached 10$^{-3}$ 
for galaxies of stellar mass $\sim$10$^{8}$ M$_{\odot}$. In other words, typically only one such low-mass field galaxy in a thousand is quenched. 
We note, however, that the quenched fraction may start to increase again for galaxies with stellar masses below 10$^7$ M$_{\odot}$ \citep{2026ApJ..1001L..42C}.

A possible quenching mechanism for dwarf galaxies in the field is  
cosmic web stripping. Here galaxies are ram pressure stripped of their gas as they pass through the low density gas associated with filaments of galaxies. This external process is particularly effective at quenching low mass dwarf galaxies, e.g. 
\cite{2013ApJ...763L..41B} and  \cite{2025ApJ...985...86B}. In the simulations of \cite{2025ApJ...985...86B} the quenching timescale associated with cosmic web stripping (i.e. the time  between 90\% and 50\% of the mass to assemble, t$_{90} -$ t$_{50}$), is on average 
4.3$\pm1$  Gyr. 
However, it does depend on the time spent within the filament \citep{2026arXiv260523457Z}. 

Internal quenching mechanisms for dwarf galaxies have focused  on feedback from stellar populations, such as stellar winds and supernovae (SN), and can generally be traced back to the seminal work of 
\cite{1986ApJ...303...39D}. 
Episodic SN feedback was invoked by 
\cite{2019MNRAS.486.2535D} to explain the large-sized low surface brightness (LSB) and ultra diffuse galaxies (UDGs) found in the field. \cite{2018MNRAS.478..906C} used the FIRE project to examine the effects of stellar feedback and outflows of gas on similar type galaxies. However, for both stellar winds and SN the gas is not completely expelled and so the galaxies do not fully quench. It is possible that both internal and external feedback mechanisms are required to explain the rare instances of quenched dwarf galaxies in low density environments. 
For a review of feedback and quenching focusing on dwarf galaxies see 
\cite{2022NatAs...6..647C}. 




Here we investigate the stellar population properties and history for two quiescent LSB dwarf galaxies that are located in low density environments. They were identified in the Ultrawide Survey of the Dragonfly Array by 
\cite{shen2024}. 
Both galaxies have large sizes (effective radii, \re $>$ 1 kpc) and low surface brightness ($\mu_{0}$ $>$ 23 mag arcsec$^{-2}$), and hence have similar properties to UDGs (see review by \citealt{Gannon2026}). 
One galaxy hosts several globular cluster (GC) candidates. We present radial age and metallicity gradients and, for the galaxy with several GCs, we directly compare the stellar populations of the GCs with the host galaxy stars. We discuss their radial gradients and star formation history in the context of possible quenching mechanisms for these non-cluster dwarf galaxies.

\section{{Data}}

\begin{figure*}
    \centering
    \includegraphics[width=\linewidth]{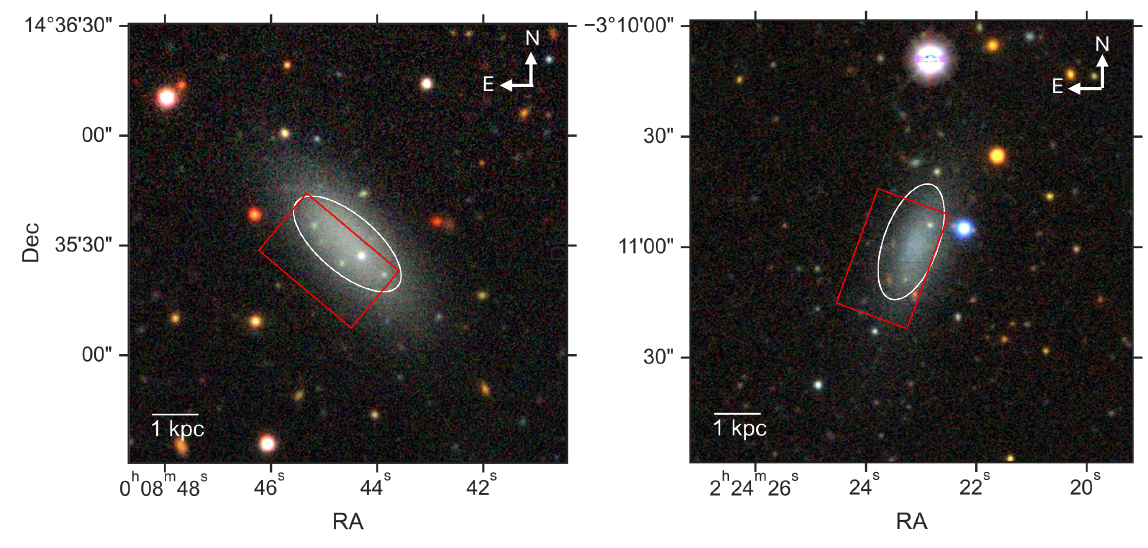}
    \caption{DFUWS~68 (\textit{left}) and DFUWS~258 (\textit{right}). Each panel shows the DESI Legacy Survey \textit{gri} colour image for with 1x \re\ (white ellipse) overlaid. The position and orientation of KCWI slicer pointing is shown by the red rectangle. A 1 kpc size bar is indicated in the lower left of each panel. \glxA\ reveals several point sources, i.e. globular cluster candidates. }
    \label{fig:overview}
\end{figure*}

\subsection{Dragonfly Ultrawide Survey}

The two main targets for this study were selected from the first results of the Dragonfly Ultrawide Survey \citep[DFUWS;][]{shen2024} which used the Dragonfly Telephoto Array \citep{2014PASP..126...55A} 
to survey the northern sky to a 1$\sigma$ surface brightness depth of 29.5~\sbu\ on $1\arcmin \times 1\arcmin$ scales. The Dragonfly Telephoto array has a total field of view (FOV) of $2^\circ \times 2^\circ$ and a pixel scale of $2.5 \arcsec$ 
after resampling. The first data release included 3100~deg$^2$ around the south Galactic cap, with the final survey expected to cover a total area of 10~000 deg$^2$ \citep{shen2024}. 

Sources were identified by visually inspecting the Dragonfly \textit{r}-band images of the first 3100~$\rm deg^2$. LSB galaxies were targeted for potential follow up through a two stage process: (1) visual inspection of the deeper low resolution imaging from Dragonfly to identify LSB features and (2) visual inspection of the corresponding coordinates in the Legacy Survey imaging \citep{dey2019}. As the Dragonfly Telephoto Array is highly optimized for LSB detection, this method of visual inspection is highly effective for identifying LSB features, and separating high surface brightness detections. However, the low resolution of Dragonfly prevents resolving the LSB structures. By combining these data with the higher resolution imaging from the Legacy Survey, LSB galaxies can be separated from other LSB structures such as unresolved background sources \citep{vanDokkumMRF}. 

The final catalogue consisted of 314 LSB galaxy candidates which were then cross-matched with SIMBAD to obtain photometric information where available. For the candidates with no available photometric data, a second round of visual inspection was performed using the higher resolution imaging from Legacy to exclude sources that were clumpy, too bright, or too small. Prioritizing galaxies with smooth and featureless morphologies targeted systems with a higher likelihood of being gas-poor and quenched. Finally, \citet{shen2024} used the \textit{g}-band Legacy Survey images to fit a S\'ersic model to the remaining 56 candidates. Spectroscopic follow-up prioritized targets with LSB ($>24$~\sbu), sizes greater than 15\arcsec, and largely isolated with no massive galaxy neighbours. In total, \citet{shen2024} reported the discovery of 11 diffuse and large LSB galaxies for which they obtained spectroscopic redshifts. 

\subsection{Target Selection}
\begin{table*}
    \centering
    \begin{tabular}{l c c c c c c c c c c c}
    \hline
        Target & RA & DEC & Velocity & $M_g$ & Dist & M$_{\ast}$ & $\mu_0(g)$ & $g-r$ & \re\ & \re\ & Environment \\ 
               & [J2000]   & [J2000] & [km~s$^{-1}$] & [mag] & [Mpc] & [M$_\odot$] & \sbu\ & [mag] & [\arcsec] & [kpc] &  \\ \hline
        DFUWS 68 & 00:08:44.56 & +14:35:30.45 & 1121 & $-$14.19 & 14.9 & 8.05 & 23.46 & 0.68 & 18.0 & 1.30 & NGC 7814 \\ 
        DFUWS 258 & 02:24:23.18 & $-$03:10:58.55 & 1333 & $-$13.18 & 15.4 & 7.64 & 23.83 & 0.54 & 16.5  & 1.23 & Field \\ \hline
    \end{tabular}
    \caption{Properties of the two galaxies presented in this study from  \citet{shen2024}, except 
    stellar mass where we assume  M$_{g,\odot}$ = +5.17 (Vega) and M/L$_{g}$ = 2. 
    }
\label{tab:targets}
\end{table*}

We assigned priorities to all 11 targets with spectroscopic redshifts from \citet{shen2024} that were observable from the Keck II telescope during the 2025B semester. The highest priority was given to targets that were classified as isolated or on the edge of a group, and that showed signs of recent quenching from the preliminary analysis done by \citet{shen2024}. Further priority was given to targets with $\mu_0(g) > 23$~\sbu\ and \re~$> 1.3$~kpc.

This resulted in two galaxies for follow up observations, DFUWS~68 and DFUWS~258, with the Keck Cosmic Web Imager (KCWI) on the Keck II telescope. Both galaxies are classified as NUDGes \citep{2025MNRAS.536.2536B}
as their surface brightness and size ``nudge'' up against the classical definition for UDGs from
\citet{2015ApJ...804L..26V}. 
Table \ref{tab:targets} gives various parameters for both targets. 


DFUWS~68 was classified by \citet{shen2024} as an LSB dwarf galaxy on the outskirts of the NGC~7814 group at a distance of 14.9~Mpc. \citet{shen2024} obtained preliminary measurements for the radial velocity, mean stellar age, and metallicity using short exposures with KCWI. 
They reported a radial velocity of $1121 \pm 7$~\kms, 
and a light-weighted metallicity of $\rm [M/H] = -0.88 \pm 0.3$. 
DFUWS~68 was assigned membership in the NGC~7814 group as it has a relative velocity of 70~\kms,  which is within 2$\times$ the velocity dispersion of the group ($\sigma = 92$~\kms). The projected separation of DFUWS~68 from the group centre is 460~kpc and is within 1.15 times the second turnaround radius ($R_{2t}$; an observational proxy for  the virial radius of the group), which was determined to be 315~kpc. However, of the 8 known dwarf galaxies within the NGC~7814 group, DFUWS~68 is the farthest from the centre of the group. 

DFUWS~258 is located in the field, with the nearest group being NGC~936 at a projected distance of 580~kpc. The closest massive galaxy in the vicinity of DFUWS~258 is UGC~1862 which has a relative velocity of 55~\kms\ and a projected separation of 270~kpc. For both \glxB\ and \glxA\ their exposure times from \citet{shen2024} were less than 40 min.



\subsection{Keck Observations}
\begin{figure}
    \centering
    \includegraphics[width=1\linewidth]{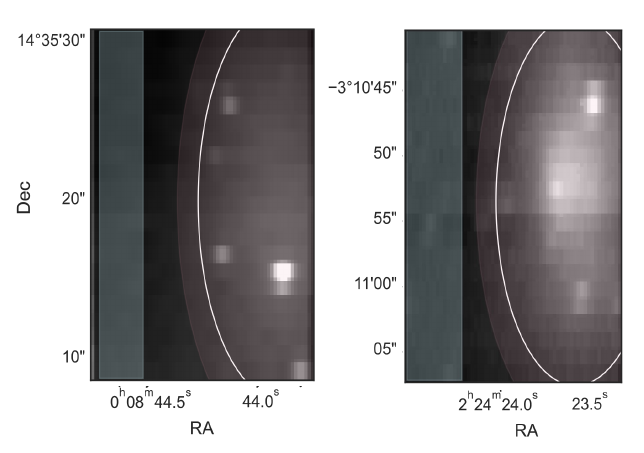}
    \caption{KCWI white light images for DFUWS~68 (\textit{left}) and DFUWS~258 (\textit{right}) with the same 1x \re\ ellipse from Figure \ref{fig:overview} overlaid. Shaded regions were used for on-chip sky subtraction. A spectrum was extracted for each galaxy up to $1.25\times$ \re.
    }
    \label{fig:whitelight}
\end{figure}

Using the integral field unit on the Keck II telescope, the Keck Cosmic Web Imager (KCWI; \citealt{2018ApJ...864...93M}), we present new  
spectroscopic data for both galaxies in this study. Observations took place on 2025 October 14  and 2025 October 15 (Program ID: W019; PI: Forbes).  We utilised both the blue and red arms available with KCWI, with both configurations using the large slicer and a full field of view of $33\arcsec \times 20.4\arcsec$. The blue arm used the BL grating with a central wavelength of 4550~\AA, giving a total wavelength range of 3600~\AA\ -- 5500~\AA,~ which covers the key stellar population absorption lines. 
The RH2 grating was used on the red arm, centred at 6585~\AA. The red side suffers from extensive cosmic rays and is not used for our stellar population analysis. 
The data were taken using $2\times2$~CCD binning. The FWHM instrumental spectral resolution of the BL and RH2 gratings is approximately 5\AA\ and 2\AA, respectively. 
Figure \ref{fig:overview} shows the position and orientation of the large slicer on the optical colour images from the Legacy Survey. Conditions for both nights were clear, with seeing ranging from 0.6\arcsec -- 1.5\arcsec. Data from the two nights were combined for a total time of 4.5\,hr on each target. Figure \ref{fig:whitelight} shows the final white-light images from KCWI. A summary of the number of exposures and integration time is given in Table \ref{tab:exposure}. \\

\begin{table}
    \centering
    \begin{tabular}{l|c|c|c}
    \hline
    Target     &  Date         &  BL/4550          & RH2/6590  \\
    \hline 
    DFUWS 68   &  2025 Oct 14  &  $5\times 1800$s  & $15\times 600$s \\
               &  2025 Oct 15  &  $4\times 1800$s  & $12\times 600$s \\
    DFUWS 258  &  2025 Oct 14  &  $4\times 1800$s  & $12\times 600$s \\
               &  2025 Oct 15  &  $5\times 1800$s  & $15\times 600$s \\
    \hline
    \end{tabular}
    \caption{A summary of the observations taken with KCWI on the Keck II telescope. The observation date and exposure times for each grating are given.}
    \label{tab:exposure}
\end{table}

These data were reduced using the standard KCWI data reduction pipeline\footnote{\url{https://kcwi-drp.readthedocs.io/en/latest/}}, excluding the automated sky-subtraction procedure. Additional post-processing of the data included trimming and corrections for flat-fielding following the procedures described in 
\cite{Gannon2020}. Finally, the data were mosaicked and stacked using \textsc{pymontage} 
\citep{2010arXiv1005.4454J}. 
We limit our spectroscopic analysis to the region within 1.25~\re\ of the target centre defined by the photometric analysis performed by \citet{shen2024}, as the S/N increases in the fainter outskirts of the galaxies. We extract spectra by collapsing all spaxels within the targeted region and isolating the wavelength range to 3700\AA~ to 5500\AA~ 
(which is the usable range common to all of the KCWI image slicers with our instrumental configuration).

To measure variations in the stellar populations as a function of radial distance, we also extract spectra from within five radial bins,  extending from the target centre at intervals of 0.25~\re. We note that we only use 4 radial bins for \glxB, combining the two outermost bins from 0.75~\re\ to 1.25~\re, to ensure S/N $>$10 in each of the bins. 
The spectra in each radial bin, and their S/N ratio, for both galaxies are shown in Figures \ref{fig:bins68} and \ref{fig:bins258}. 
Moreover, all other sources within the extracted region were masked (see Fig. \ref{fig:overview}). We provide a more detailed description of the point source detection and analysis in Section \ref{sec:globs}, however qualitatively \glxA\  appears to be richer in GCs than \glxB. 

\begin{figure}
    \centering
    \includegraphics[width=1\linewidth]{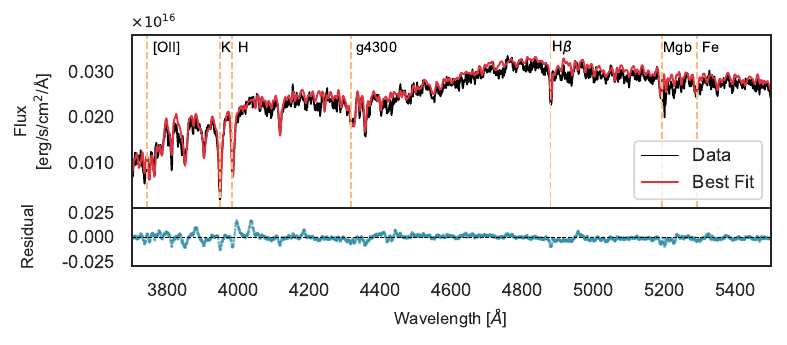}
    \includegraphics[width=1\linewidth]{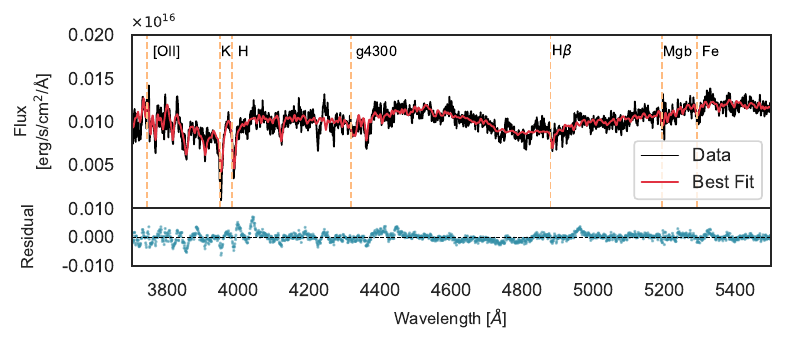}
    \caption{Integrated KCWI spectra extracted within $1.25\times$~\re\ for DFUWS~68 (\textit{top}) and DFUWS~258 (\textit{bottom}) in black. Both spectra have S/N $\sim$ 30. The best fit from \textsc{pPXF} for each spectrum is shown in red. The residual between the best fit and the data is shown in aqua green in the subpanel. The key spectral lines are indicated by vertical dashed orange lines. 
    }
    \label{fig:ppxf_fit}
\end{figure}


\section{Stellar Populations}

\subsection{Spectral Analysis }


The integrated and aperture spectra for each galaxy were fitted using the full-spectrum-fitting routine \citep[\textsc{pPXF;}][]{cappellari2004, cappellari2017, cappellari2023} to extract recessional velocity and stellar population information. We utilise the E-MILES stellar templates \citep{vazdekis2015} with a Kroupa universal initial mass function \citep{kroupa2001} and the BaSTI (a Bag of Stellar Tracks and Isochrones) isochrones \citep{higalgo2018}. This provides fine sampling of both metallicities ($\rm -2.27 \leq [M/H] \leq 0.40~dex$) and ages (30~Myr \mbox{--} 14~Gyr), which has been shown to be appropriate to obtain the stellar populations of low-mass galaxies 
(\citealt{ferremateu2025};
\citealt{Levitskiy2025}). 
This choice of templates allows for a wider range of metallicities than the \textsc{FSPS} SSP template spectra \citep{conroy2009, conroy2010} used in \citet{shen2024}, particularly to lower metallicities. %
We note that our chosen combination of stellar libraries provides a small update to the methods presented in \citet{shen2024} who used the \textsc{FSPS} SSP template spectra \citep{conroy2009, conroy2010}, based on the MILES stellar library \citep{sanchez2006}, restricting the metallicity range to $\rm -1.5 \leq [M/H] \leq 0.50~dex$ while allowing the age to vary across the whole range of 1--17.8~Gyr. For further details of the spectral analysis see \cite{ferremateu2025}. 

We use a 15th-order Legendre multiplicative polynomial and a 1st-order additive polynomial (essentially a pedestal for the continuum) when fitting. These help account for any offsets in the continuum shape relative to the templates that may be introduced due to an imperfect flux calibration. 
We verify the minimal effect of the exact choice of polynomial by order repeating the fitting procedures with multiplicative polynomials of orders 5 \mbox{--} 30, confirming minimal variations within the measured uncertainties. 
We use the median solutions of 1000 bootstrap fitting iterations with 1$\sigma$ uncertainties taken as the 16th and 84th percentiles. These are the formal uncertainties from the fitting process and do not include systematic uncertainties due to a different fitting process or the use of different stellar population libraries etc. 
Our stellar population results are summarised in Table~\ref{tab:SSP}. 
Throughout this work, we report the mass-weighted stellar population results (light-weighted results can be found in Appendix 
\ref{app:light-weighted}). 

Figure \ref{fig:ppxf_fit} shows the continuum-corrected integrated spectrum for each galaxy with the 
best fit over-plotted. Both spectra have a signal-to-noise (S/N) of $\sim$30. 


\begin{table*}
    \centering
    \caption{Table of stellar population results.}
    \begin{tabular}{l l c c c c c c c c c}
    \hline
        Target & Region & [M/H] & Age & $t_{50}$ & $t_{90}$ & $t_q$ \\ 
         &  & [dex] & [Gyr] & [Gyr] & [Gyr] & [Gyr] \\ \hline
        DFUWS 68 & galaxy & --0.90$^{+0.10}_{-0.06}$ & 8.70$^{+0.94}_{-0.89}$ & 5.0& 5.5 & 0.5 & \\ 
        ~ & 68e & $-1.22^{+0.07}_{-0.08}$ & $8.20^{+0.98}_{-0.97}$ & 5.5 & 8.0 & 2.5\\ 
        ~ & GCs & $-0.94^{+0.13}_{-0.09}$ & $9.23^{+1.20}_{-1.35}$ & 5.0 & 6.0 & 1.0 & \\ 
        ~ & 0--0.25~\re\ & $-0.95^{+0.06}_{-0.06}$ & $8.75^{+0.84}_{-0.76}$ & $5.0$ & $6.0$ & 1.0 \\ 
        ~ & 0.25--0.50~\re\ & $-0.92^{+0.07}_{-0.05}$ & $8.65^{+0.84}_{-0.72}$ & $5.0$ & $6.0$ & 1.0 \\ 
        ~ & 0.50--0.75~\re\ & $-0.93^{+0.08}_{-0.05}$ & $8.92^{+0.88}_{-0.81}$ & $5.0$ & $6.0$ & 1.0  \\ 
        ~ & 0.75--1.0~\re\ & $-0.89^{+0.09}_{-0.07}$ & $8.53^{+0.98}_{-1.01}$ & $5.0$ & $6.0$ & 1.0  \\ 
        ~ & 1.0--1.25~\re\ & $-0.86^{+0.12}_{-0.09}$ & $9.05^{+1.30}_{-1.31}$ & $5.0$ & $6.0$ & 1.0  \\ 
        DFUWS 258 & galaxy & --1.09$^{+0.14}_{-0.09}$  & 9.01
        $^{+1.24}_{-1.16}$ 
        & $5.0$ & $6.5$& 1.5\\ 
        ~ & 0--0.25~\re\ & $-1.15^{+0.10}_{-0.12}$ & $7.45^{+1.01}_{-1.11}$ & $6.0$ & $8.5$ & 2.5  \\ 
        ~ & 0.25--0.50~\re\ & $-1.11^{+0.10}_{-0.08}$ & $8.73^{+1.06}_{-1.29}$ & $5.5$ & $6.5$ &  1.0  \\ 
        ~ & 0.50--0.75~\re\ & $-0.94^{+0.14}_{-0.07}$ & $9.63^{+0.89}_{-1.36}$ & $5.0$ & $5.5$ & 0.5  \\ 
        ~ & 0.75--1.25~\re\ & $-1.00^{+0.19}_{-0.13}$ & $8.74^{+1.28}_{-1.17}$ & $5.0$ & $6.5$ & 1.5  \\ 
        \hline
    \end{tabular}
    \label{tab:SSP}
\end{table*}

\begin{figure}
    \centering
    \includegraphics[width=1\linewidth]{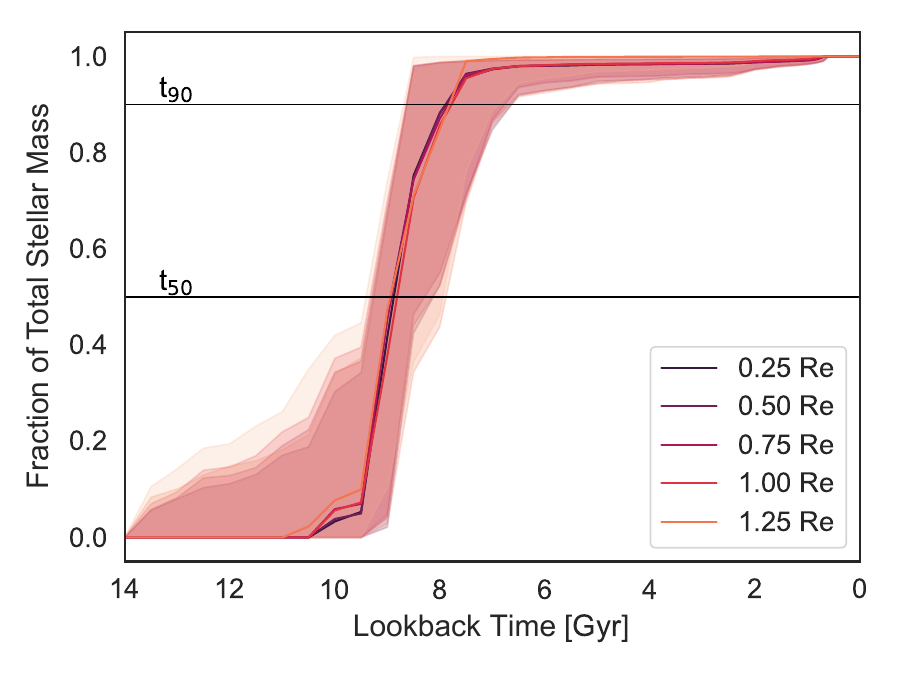}
    \includegraphics[width=\linewidth]{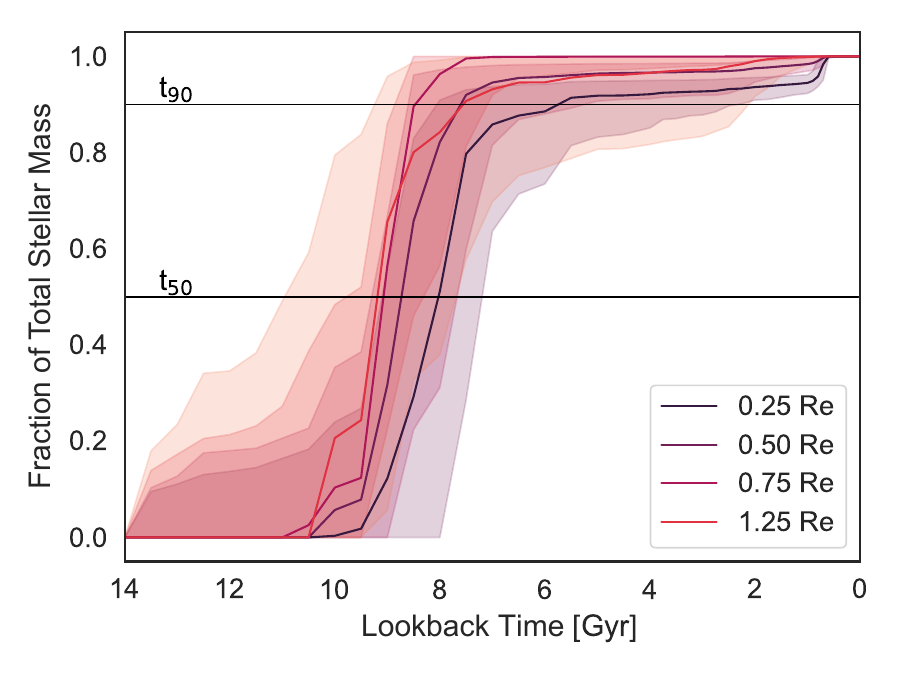}
    \caption{Star formation histories (SFHs) as a function of lookback time, 
    for each radial bin  as described in the legend. The radial bins reach $1.25\times$\re\ for DFUWS~68 (\textit{top}) and DFUWS~258 (\textit{bottom}). 
    The SFH for \glxA\ (\textit{top}) shows uniform quenching timescales at all radii. 
    The SFH for \glxB\ (\textit{bottom}) shows the inner regions continuing to form stars for $\sim 2\mbox{--}3$~Gyr after the outer regions reach the $t_{90}$ threshold. 
    DFUWS~258's SFH shows signs of a recent burst of star formation within the past $\sim$2~Gyr, however it does not appear to contribute significantly to the overall mass of the system.
    }
    \label{fig:SFH}
\end{figure}

\subsection{Stellar Populations and Star Formation Histories}

We find that overall both galaxies have old ages and low metallicities, i.e. mass-weighted ages of $8.70$~Gyr and $9.01$~Gyr  and metallicity  [M/H]$=-0.90$ and $-1.09$ for \glxA\ and \glxB, respectively. These metallicities are consistent with the local dwarf galaxy mass--metallicity relation and its scatter as presented by \cite{2019ARA&A..57..375S}.

Figure \ref{fig:SFH} shows the derived star formation histories (SFHs) for all apertures of \glxA\ (\textit{top}) and \glxB\ (\textit{bottom}). 
In Table~\ref{tab:SSP} we list the 
time to reach 50\% and 90\% of the total mass, and the quenching timescale (t$_{90}$ -- t$_{50}$). 
Typical uncertainties are about a Gyr. 
We also include the global mass-weighted age and metallicity. These parameters are given for each radial bin and, in the case of \glxA, the stacked GCs and the possible nuclear star cluster (source 68e; see section 5.3). 

We find that both galaxies formed on rather  similar timescales. 
In particular, \glxA\ shows virtually the same quenching timescales at all radii,  occurring on average early ($t_{90} = 5.5~\rm Gyr$) and quickly 
($t_{q} = 0.5$ Gyr).  
The SFH for \glxB\ 
shows more varied formation timescales, 
with the inner regions continuing to form stars for $\sim 2\mbox{--}3$~Gyr after the outer regions reach 90\% of their stellar mass. While $t_{50}$ is similar in both cases, the quenching occurs on average slower in \glxB\  ($t_q = 0.5\mbox{--}2.5~\rm Gyr$). Additionally, 
DFUWS~258 shows signs of some recent star formation within the past $\sim$2~Gyr, although it does not contribute significantly to the overall mass of the system.


\begin{figure*}
    \centering
    \begin{subfigure}{0.45\linewidth}
        \includegraphics[width=\linewidth]{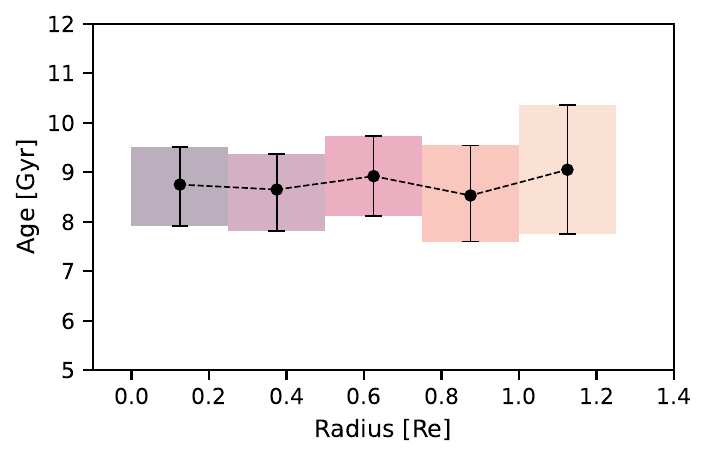}
    \end{subfigure}
    \hfill
    \begin{subfigure}{0.45\linewidth}
        \includegraphics[width=\linewidth]{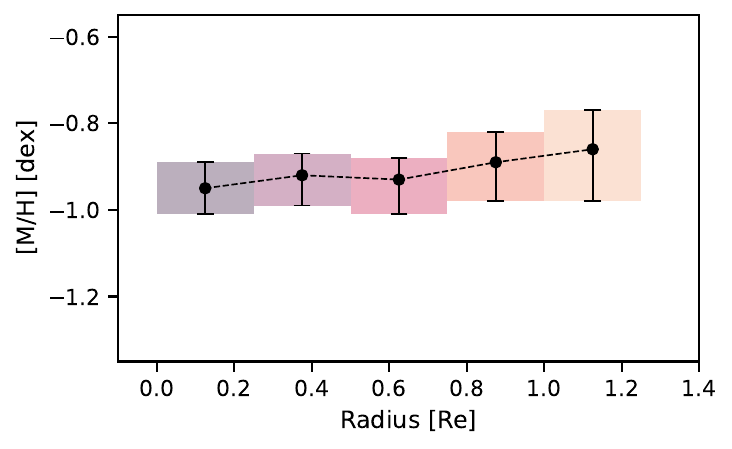}
    \end{subfigure}

    \begin{subfigure}{0.45\linewidth}
        \includegraphics[width=\linewidth]{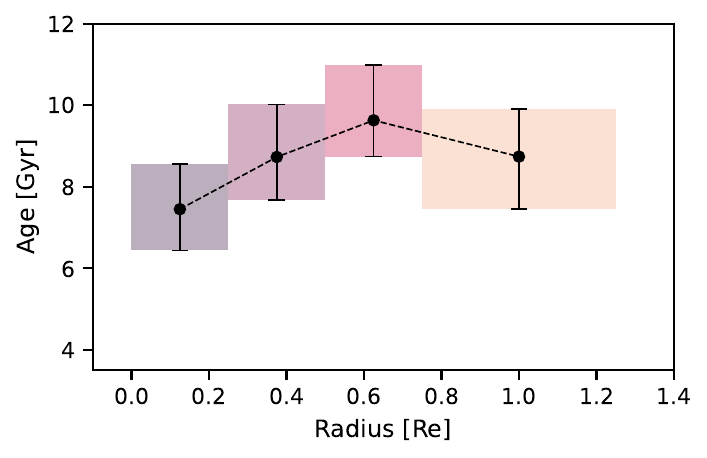}
    \end{subfigure}
    \hfill
    \begin{subfigure}{0.45\linewidth}
        \includegraphics[width=\linewidth]{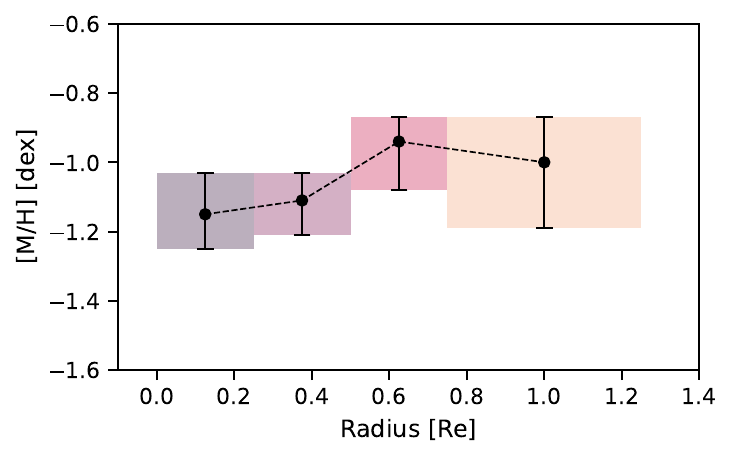}
    \end{subfigure}
    
    \label{fig:mw_gradients}

    \caption{Age and metallicity profiles along radial bins extending to $1.25\times$~\re. For each bin, the median value is shown with upper and lower bounds indicating the 16th and 84th percentile for the bootstrapped results. The shaded colours match the colours used to differentiate the different radial bins in Figure \ref{fig:SFH}. (\textit{top row}:) The median age (\textit{left)} and metallicity (\textit{right)} for DFUWS~68. Each radial bin steps outward in increments of 0.25~\re\ to a maximum radius of $1.25\times$ \re\ from the galaxy centre. 
    {\it (bottom row:)} The median age (\textit{left)} and metallicity (\textit{right)} DFUWS~258. Similar to the top panel, the radial bins step outward increments of 0.25~\re\ with the exception of the outermost bin which combines the outer region from 0.75\mbox{--}1.25~\re. DFUWS~68 shows near-flat age and metallicity gradients. DFUWS~258 shows signs of a flat-to-positive age gradient and a positive metallicity gradient.
    }
   \label{fig:stell_pop}
\end{figure*}


\subsection{Age and Metallicity Radial Gradients}

Recent studies have uncovered growing evidence for flat-to-positive metallicity gradients, alongside flat age profiles, within a limited sample of UDGs with spatially resolved data \citep{fensch2019, villaume2022, ferremateu2025, buzzo2025_fcc, Levitskiy2025}. This deviates from what is observed in classical dwarf galaxies \citep[e.g.][]{chilingarian2009, koleva2011, sybilska2017, taibi2022}, as well as those produced for both simulated classical dwarfs and UDGs \citep[e.g. NIHAO and TNG50;][]{cardona2023, benavides2024}. 

We 
measure the spatial variations in the stellar population parameters, for  several different apertures to a radius of 
$1.25\times$\re\, in  Figure \ref{fig:stell_pop}. The panels show the derived median mass-weighted ages and metallicities for each of the bins as a function of effective radius for both galaxies. We obtain the radial gradient by normalizing within 1~\re. We find virtually flat age profiles for both \glxA\ and \glxB ($\rm \nabla \log Age$  = 0.01 and 0.08\;log(Gyr/R/$R_\mathrm{e}$) respectively), similar to those previously measured in other UDGs \citep[e.g.][]{villaume2022, ferremateu2025, Levitskiy2025}. We also find flat-to-rising metallicity gradients in both targets ($\rm \nabla \log [M/H]$ = 0.06 and 0.22\;dex/log(R/$R_\mathrm{e}$)). 
These results follow the trends observed in the small number of other (mostly cluster) UDGs with spatially resolved stellar populations, despite the relative isolation of these two galaxies.

We show the light-weighted stellar population gradients in Appendix \ref{app:light-weighted} which better trace any signature of recent stellar activity within our targets. For \glxB, we find a more pronounced positive age gradient, indicating a recent burst of star formation concentrated within the central regions of the galaxy. This is corroborated by the cumulative luminosity profiles in Figure \ref{fig:sfh_lw}, which reveals a 20\% increase in total luminosity within the innermost annular bin, despite this representing only a negligible fraction of the total mass-weighted stellar population. Conversely, \glxA ~exhibits consistently near-flat age and metallicity gradients across both light-weighted and mass-weighted measures, aligning with the global profiles discussed above.

\section{Globular Clusters associated with DFUWS 68} \label{sec:globs}

Using optical imaging from the Legacy Survey \citep{dey2019}, both \glxA\ and \glxB\ reveal point sources that may be GCs (see Fig.~\ref{fig:legacy}). 
\glxA\, in particular, appears to  host several GCs, including one that is close to the galaxy centre.  Below, we describe the identification of point sources in our KCWI data for \glxA\ and the extraction of spectra from our KCWI data. We were unable to obtain a spectrum for any point sources in \glxB.


    \label{fig:legacy}

\subsection{GC Candidate Selection}

In order to better distinguish GC candidates from the host galaxy we follow a method similar to \citet{shen2024}.
First we create a model galaxy light profile using the simple single-component S\'ersic model from \textsc{Astrophot} 
\citep{2023MNRAS.525.6377S}. 
We utilise all three optical bands (\textit{gri}) available from the Legacy Survey to create light profiles for each galaxy. Initial estimates for the size, central surface brightness, and S\'ersic index were chosen based on the fit parameters from \citet{shen2024}, however, we allowed each to be fit as a free parameter to account for any minor offsets between the different optical bands. We note that all of the S\'ersic model parameters matched those presented in \citet{shen2024} within uncertainties. 

The modelled galaxy light profile was then subtracted from the image cutout for each of the respective bands. The resulting residual image uncovers any underlying bright sources within the galaxy. The Legacy survey image, model created and residual image for the g-band are shown in 
Fig.~\ref{fig:legacy}. The five point sources identified within 1~R$_e$ are labelled a-e. 
We measure their total magnitudes from the residual image using 
\textsc{SExtractor} 
\citep{1996A&AS..117..393B}.

\subsection{GC Spectroscopic Analysis}

Spectra the 5 point sources were extracted from the datacube using an elliptical aperture appropriate for each GC relative to its position on the image slicer. 
A local annulus around each source is then used to remove the galaxy  background light. 
We follow the same full-spectrum fitting methods outlined above to obtain radial velocities for each of the GC candidates from the KCWI data. 
While photometric analysis revealed 5 GC candidates around \glxA, 
we were unable to obtain a spectrum for 68b and do not discuss this source further. For the 4 remaining GC candidates all have radial velocities indicating that they are associated with the host galaxy and therefore likely bona fide GCs.

\subsection{Measured GC properties}

In Table~\ref{tab:gc_prop} we summarise their properties 
including the difference in velocity with respect to \glxA, the spatial offset from the galaxy centre, the GC colour, magnitude and inferred stellar mass. 
We estimate stellar masses assuming M$_{g,\odot}$ = +5.17 (Vega) and M/L$_{g}$ = 2. 
We note that all of the GC candidates inhabit a narrow colour range with little deviation from the measured colour of their host galaxy (i.e. $g-r = 0.68$). 


\begin{table}
    \centering
    \begin{tabular}{c|c|c|c|c|c|c}
    \hline
       Source & $\Delta$RV & offset & \textit{g$-$r} & $m_g$  & $\rm \log~M_*$ \\
              & [\kms] &  [$\rm R/R_e$] & [mag] & [mag]  & [$\rm M_\odot$]\\
       \hline
       DFUWS~68a & $-52 \pm 13$ & 0.62 & 0.53 & 22.65 & 5.66\\
       DFUWS~68c & $-24 \pm 13$ & 0.32 & 0.63 & 23.24 & 5.42\\
       DFUWS~68d & $3 \pm 16$ & 0.68 & 0.77 & 23.46 & 5.33\\
       DFUWS~68e & $7 \pm 5$ & 0.24 & 0.62 & 21.51 & 6.11\\
       \hline
    \end{tabular}
    \caption{Globular cluster parameters. For each GC the table lists velocity relative to the host galaxy, the projected radius, g--r colour, g band magnitude and inferred stellar mass. We were not able to measure a radial velocity for source 68b.}
    \label{tab:gc_prop}
\end{table}


Combining the stellar mass all 4 GCs gives a total of 2.2 $\times$ 10$^{6}$ M$_{\sun}$ or 2\% of the stellar mass of the galaxy. This is a lower limit given the GC system likely contains many more GCs, although these GCs are perhaps the brightest and most massive in the GC system, dominating the total mass. 
As discussed by \cite{2025MNRAS.536.1217F} 
this percentage is much higher than classical dwarf galaxies but is similar to that measured for UDGs -- particularly quenched UDGs that lie in high density environments.  
This, combined with the quenched nature of \glxA,  resembles that of GC-rich ``failed galaxy" 
(\citealt{Danieli2022}; \citealt{2025MNRAS.536.1217F}). 


\subsection{Comparing Galaxy and GC Stellar Populations}


\begin{figure}
    \centering
    \includegraphics[width=1\linewidth]{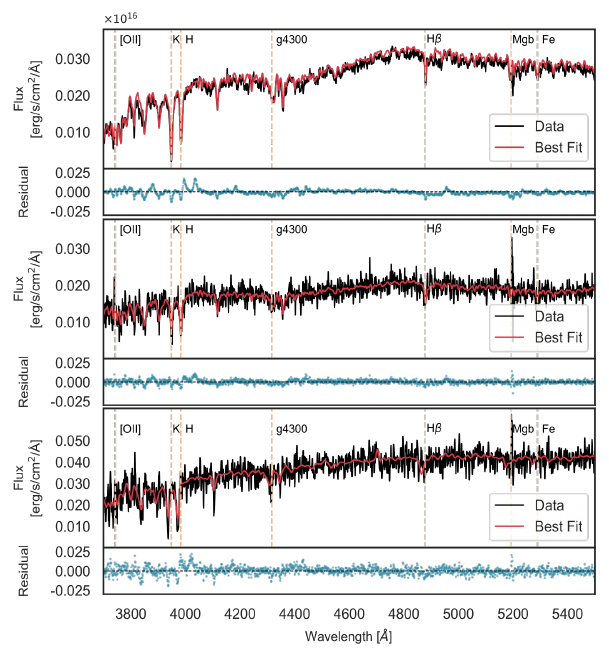}
    \caption{KCWI spectra of the galaxy \glxA\ \textit{(upper}), the brightest GC 68e  \textit{(middle)}, combined globular clusters \textit{(bottom)}. The S/N ratios are $\sim$30, 12 and 12 respectively. 
    Data are shown by the black line, the fit by the red and residuals to the fit by the aqua green line in the subpanel.  Common spectral lines are labelled. The galaxy main body, combined GCs and brightest GC reveal similar stellar populations.
    }
    \label{fig:D68_comb0_spec}
\end{figure}

\begin{figure}
    \centering
    \includegraphics[width=\linewidth]{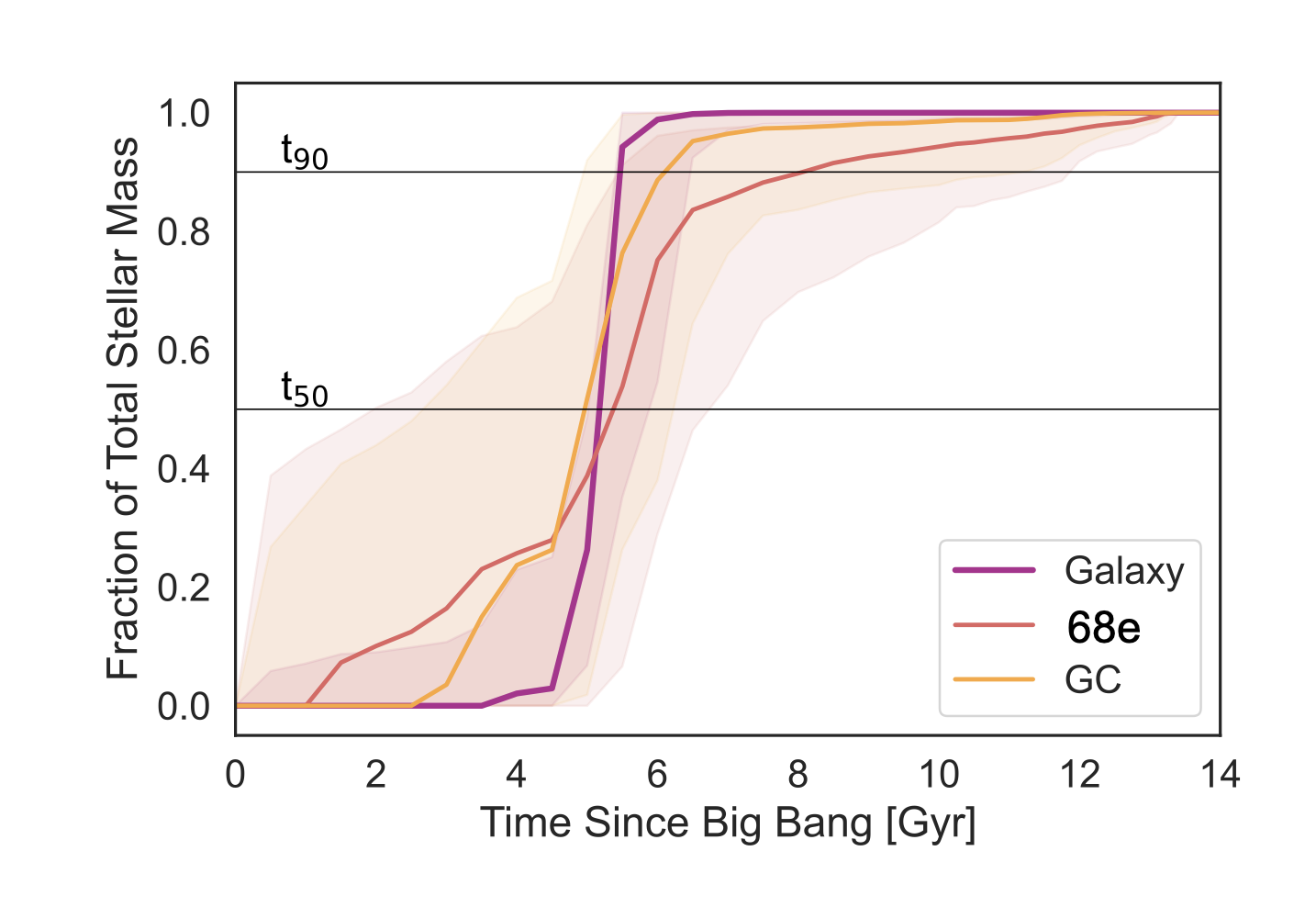}
    \caption{Star formation history for the possible nuclear star cluster (68e), the combined GCs and the host galaxy \glxA.
    Shaded regions show the uncertainty. Horizontal lines represent the time to acquire 50\% and 90\% of the total stellar mass of the galaxy. All three have similar star formation histories and timescales. 
    }
    \label{fig:gc_nsc_sfh}
\end{figure}

We combine and stack the spectra for the \glxA\ GCs a, c and d to improve the S/N (to $\sim12)$. Source \glxA b is excluded due to its non-detection in the KCWI data, whereas DFUWS~68e, the brightest source, has a S/N of 12 allowing for its own  separate analysis. 
We follow the same full-spectrum fitting procedure used for the galaxy spectra to measure the stellar populations of the GCs. 
The spectra of the combined GCs, the brightest GC and the host galaxy \glxA\ are shown in Fig.~\ref{fig:D68_comb0_spec}.
We find that the average mass-weighted age and metallicity of the combined GC spectra are approximately similar to those found from the global galaxy spectrum. In particular, we find old ages of around 9 Gyr and low metallicity of [M/H] $\sim$ --1. 
Within the uncertainties, we find similar old ages and 
metallicity for source 68e. Thus the GCs and galaxy stars formed at roughly the same time from the same enriched gas. 
The (rapid) quenching timescales are also similar for the galaxy and GCs. Figure \ref{fig:gc_nsc_sfh} shows the measured star formation histories for each of these components.
Each shows a similar shaped SFH. The t$_{90}$, t$_{50}$ ages and quenching timescales are listed in Table \ref{tab:SSP}. 

\section{Discussion}



\subsection{Flat-to-Rising metallicity gradients}

\begin{figure}
    \centering
    \includegraphics[width=\linewidth]{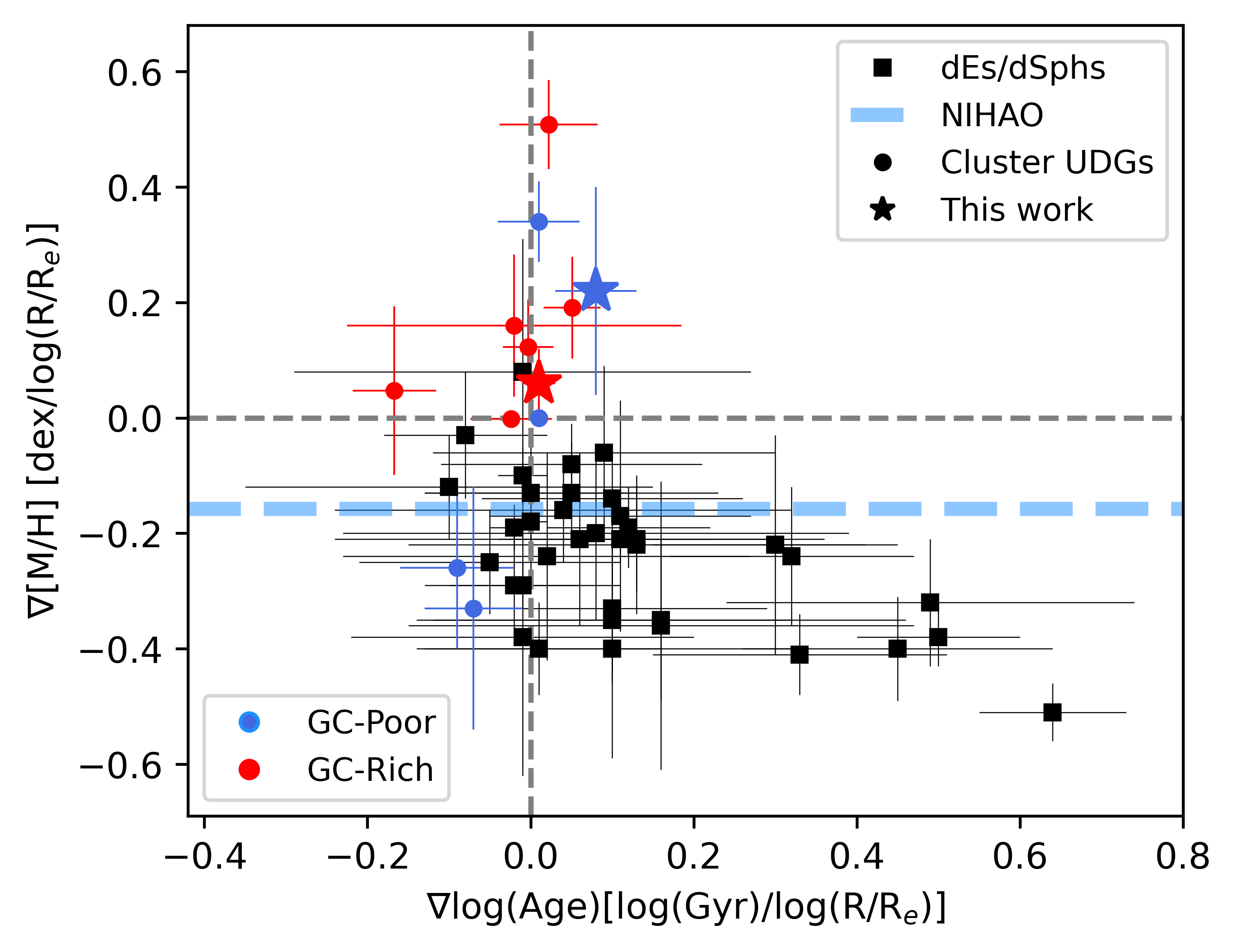}
    \caption{Age and Metallicity gradients. Measured age vs metallicity gradients for \glxA\ (red star) and \glxB\ (blue star). As per the legend, black squares show classical dwarf galaxies in clusters, black circles for UDGs located in clusters, and the mean metallicity gradient predicted by NIHAO simulations for field LSB galaxies (age gradients are not available) \glxA\ and \glxB\ show flat-to-rising metallicity gradients similar to UDGs, and unlike the negative gradients of classical dwarfs. }
    \label{fig:delta}
\end{figure}

In Fig.~\ref{fig:delta} we show the measured gradients for \glxA\ and \glxB\ compared to classical dwarf galaxies (dE/dSph) located mostly in clusters from the literature 
(i.e. \citealt{2011MNRAS.417.1643K}; \citealt{sybilska2017}; \citealt{2022MNRAS.515.4622B}) plus a sample of UDGs (and NUDGes) from 
\cite{ferremateu2025} and \cite{Levitskiy2025}. The UDG sample is coded by GC system richness, with richness being defined as more than 20 GCs. Until deeper multifilter, or space-based, imaging becomes available we assign qualitative GC richness to \glxA\ (GC-rich) and \glxB\ (GC-poor). 
Here \glxA\ and \glxB\ have metallicity gradients similar to UDGs and fairly distinct from those typically found for classical dwarfs. While there is a weak trend for more positive metallicity gradients in GC-rich UDGs, this is less clear for the two galaxies in this study (the more positive gradient galaxy, \glxB, has fewer GCs).  
It is of course only two galaxies and a larger sample is required to determine whether they follow the GC richness trend or not.

Fig.~\ref{fig:delta} also shows the predictions of the NIHAO simulation of dwarf galaxies (UDGS and NUDGes) located in low density environments similar to \glxA\ and \glxB. In these models the main source of feedback is from SN. The work of 
\cite{cardona2023} predicted mean metallicity gradients (but no age gradients) of --0.18 dex/log(R/R$_e$), with a full range of --0.5 dex/log(R/R$_e$) to zero gradient. They also found larger-sized and rotationally-supported galaxies to have steeper positive metallicity gradients (but none actually reached a positive gradient). We also note the Romulus simulations of isolated UDGs by 
\cite{2021MNRAS.502.5370W} which form by early mergers. They did not describe metallicity or age gradients of their model UDGs but did predict negative colour gradients, which is contrary to that observed \citep{Levitskiy2025}. 
The simulations of \cite{cardona2023}, like many others, under-predict the global metallicity for a given galaxy stellar mass. This might suggest that the internal feedback mechanism, driving outflows, in these models is too strong. Recent work on dwarf galaxies with stellar masses similar to our targets in the SAGA survey has also suggested that cosmological simulations of galaxy formation may be overestimating star-formation feedback \citep{KadoFong2025}.

\subsection{Quenching Mechanisms}

Both galaxies in this study are located in low density environments and thus are not subject to the various processes that can quench galaxies in a cluster environment. 
For \glxA\ one may want to invoke the nearby group environment as a possible cause of environmental quenching. In studying the dwarf galaxy satellites around the Milky Way it has been found that galaxies of a stellar mass greater than $10^8$~M$_\odot$~(as \glxA\ just is) are largely resistant to rapid environmental quenching \citep{Akins2021}. Due to the rapid nature of \glxA's quenching, it seems implausible for the nearby group environment to be the cause. 

We also note the work of 
\cite{2025OJAp....8E.110S} who used 
KCWI to study the stellar populations of several UDGs/NUDGes. From their total sample of 44 galaxies they identified 7 to have K+A like spectra similar to \glxA\ and \glxB.  Of the 7, 4 were satellites of a more massive galaxy, 1 may be a back-splash galaxy but 2 appeared to be quite isolated. 
They did not conduct a full spectral analysis nor measure any radial properties. Nevertheless, their two isolated quiescent galaxies have sizes, stellar masses, and quenching timescales, comparable to \glxA\ and \glxB. 
Next, we discuss the possible mechanisms for quenching \glxA\ and \glxB.\\

{\it Could these galaxies have been quenched by cosmic web stripping?}\\

The TNG50 simulations of \cite{2025ApJ...985...86B} predict a quenching 
timescale for dwarf galaxies stripped by a cosmic web of t$_{90}$ -- t$_{50}$ $\sim$ 3--6 Gyr. We measure equivalent quenching times of 0.5 and 1.5 Gyr for DFUWS 68 and 258,  respectively. Thus both galaxies have quenched more rapidly than 
expected for cosmic web stripping. 
We note that although ram pressure stripping in a cluster is more efficient and operates on a shorter timescale, of a Gyr, both galaxies lie well beyond any back-splash region of any  cluster. 
Cosmic web stripping is also expected to operate outside-in so that the outer regions will be younger and more metal-rich, as is found in most classical dwarf galaxies 
\citep{2011MNRAS.417.1643K}. However, like many UDGs (\citealt{ferremateu2025}; 
\citealt{Levitskiy2025}),  
we measure flat-to-rising age and metallicity gradients (see Fig.~\ref{fig:stell_pop}) which is more suggestive of 
an inside-out process.
\\

{\it Could these galaxies be quenched by internal processes?}\\

The quenching timescale for SN is from millions of years up to a Gyr, and so is consistent with the quenching times we measure for \glxA\ and \glxB\ (see Table 3). 
However, as noted above, simulations that incorporate SN feedback 
\citep{cardona2023} for field UDGs/NUDGes (i.e. similar to \glxA\ and \glxB), do not produce the positive metallicity gradient we find for \glxB. 
We note that 
\cite{Levitskiy2025} investigated the TNG50 simulations of 
\cite{benavides2024} which modelled quiescent UDGs located in clusters. In general, these TNG50 simulations also struggled to reproduce 
the positive metallicity gradients observed for cluster UDGs. 

While SN feedback is no doubt playing a role in quenching, it does not appear to be able to explain a key observational property, i.e. positive metallicity gradients.  
For the latter, GCs may be responsible.
For GC-rich galaxies, like \glxA, a vigorous burst of GC formation at early times may also contribute to 
quenching subsequent star formation (\citealt{Danieli2022}; \citealt{2025MNRAS.536.1217F}). As these GCs are destroyed by tidal shocks over time, they will deposit old, metal-poor stars preferentially in the central regions of the 
galaxy potentially giving rise to the observed flat-to-rising metallicity profiles 
(Levitskiy et al. 2026, submitted). 



    \subsection{Possible Off-Centre Nuclear Star Cluster}
  \label{sec:nsc}  
  
\begin{figure}
    \centering
    \includegraphics[width=0.9\linewidth]{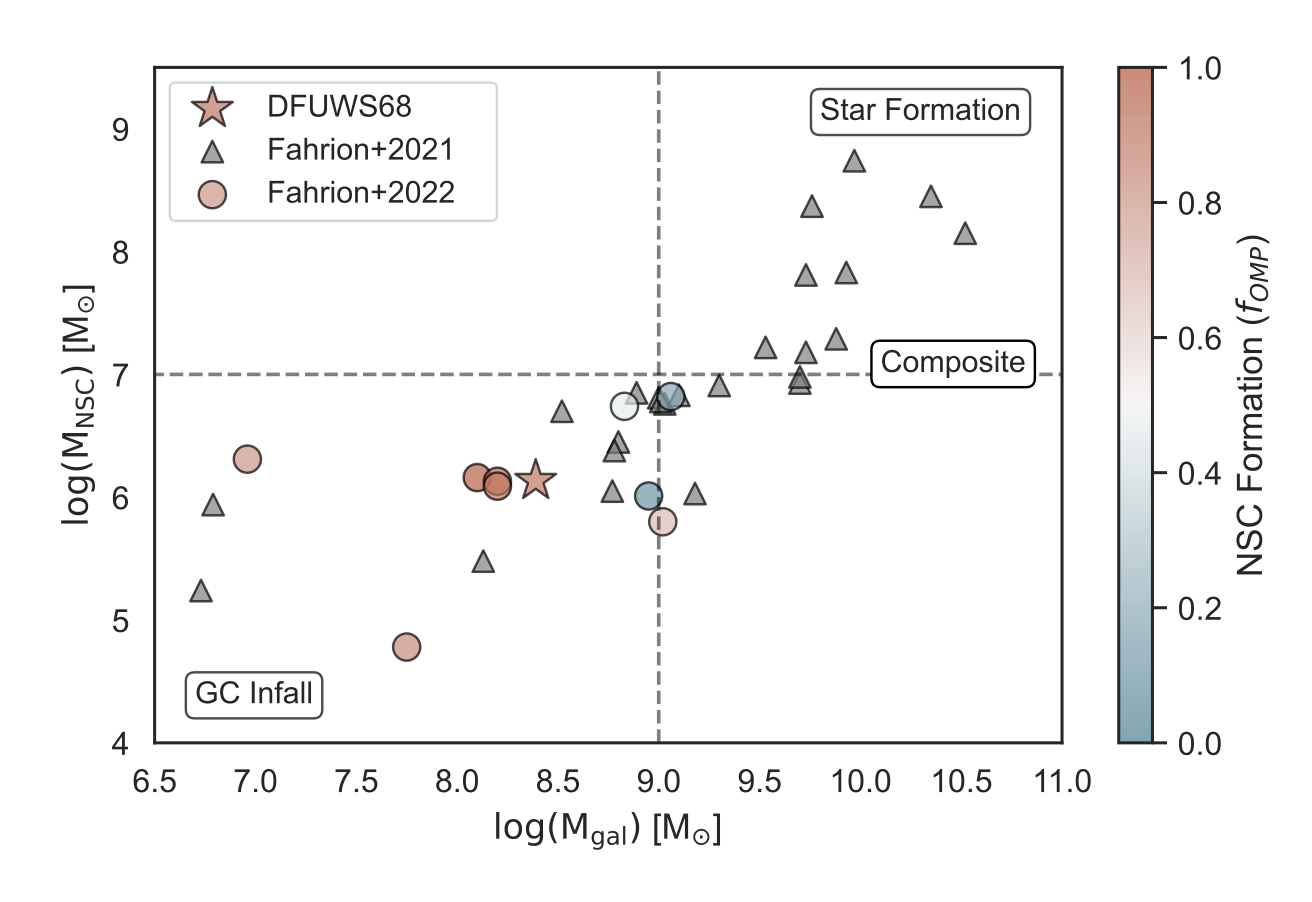}
    \caption{The masses of nuclear star clusters (NSCs) are plotted vs their host galaxy mass using literature data from \citet{Fahrion2021, Fahrion2022a, Fahrion2022b}. Data are colour-coded by the fraction of metal-poor stars within the NSC to aid the diagnostic ability of the diagram. Following \citet{Fahrion2022a}, the dominant formation mechanism for the NSC is indicated as either globular cluster (GC) infall, star formation via gas flows or a composite of the two. The plotted star symbol, denoted as \glxA\ in the legend, corresponds to source 68e (a possible off-centre nuclear star cluster). It is plotted on the NSC diagram using the stellar mass of \glxA\ (from Table 1) and the mass of 68e (from Table 4). 
    Source 68e lies within the GC infall region of the diagram. 
    }
    \label{fig:NSC_form}
\end{figure}

    The brightest point source in \glxA\ (68e), with a stellar mass of over 10$^6$ M$_{\odot}$, is located near the galaxy centre at a projected radius of 0.24 times the effective radius and with a velocity difference to the galaxy of only 7 $\pm$ 5 km s$^{-1}$.  If it migrates to the galaxy centre via dynamical friction, this source might be classified as a nuclear star cluster (NSC). Although we note that in a study of 78 dwarf galaxies with nuclei, \cite{2000A&A...359..447B} found 20\% to have off-centre nuclei with displacements up to 1~R$_e$. As discussed by \citet{Neumeyer2020}, it is commonly theorised that NSCs may have formed by star formation from gas inflows or via the mergers of several infalling GCs due to dynamical friction (or a combination of both processes). To compare source 68e in \glxA\ with known NSCs we show it on the commonly used diagnostic plot shown in Figure \ref{fig:NSC_form}. As shown by \citet{Fahrion2021, Fahrion2022a, Fahrion2022b}, this plot helps to separate out GCs that have formed via these two pathways. 

    Figure \ref{fig:NSC_form} shows the mass of the NSC compared to the host galaxy mass as colour-coded by the metal-poor fraction of the NSC. \citet{Fahrion2022a} defines this fraction to have ages $> 2~ \rm Gyr$ and metallicities $\rm [M/H] < -1.0$. The region of low mass NSCs in lower mass galaxies is dominated by high fractions of old and metal-poor stars. Due to the similarities of these stars with those in GCs, this region of the plot is associated with NSC formation via GC infall. For high mass NSCs, in higher mass galaxy hosts, the metal-poor fraction tends to be much lower, which is suggestive of recent, more metal-rich star formation. As such, this region of the plot tends to be associated with NSCs that form via star formation induced by gas flows. 
    The source 68e in \glxA\ (red star) lies squarely within the GC infall region of the diagram and with a relatively high metal-poor fraction. We conclude that it is likely a GC (or a merger of several GCs) and speculate that is sinking towards the galaxy kinematic and photometric centre due to dynamical friction. So it may be considered as a (massive) GC today and a possible off-centre NSC.  
    
    This raises an interesting question as to the nature of \glxA's dark matter halo. In general, it is expected that dynamical friction is much less efficient in cored dark matter halos due to a process known as `core stalling' \citep{Read2006, BinneyAndTremaine}. Many UDGs have been shown to have cores (see e.g., \citealp{Gannon2020, Gannon2022, Gannon2026} and \citealt{Levitskiy2025})  
    so it leaves an open question as to why \glxA's dark matter profile is perhaps not cored and thus different to other UDGs, or how dynamical friction has managed to be so efficient in a cored halo. Unfortunately, our data have insufficient spectral resolution/signal-to-noise to measure the velocity dispersion (and dynamical mass) that would be required to disentangle these questions.

    \subsection{GC System Properties}

   As summarised in \citet{Gannon2024, Gannon2026}, there are fewer than half a dozen UDGs for which the GCs and host galaxy stars both have spectroscopically measured stellar populations. Perhaps the best known is NGC5846\_UDG1 (see \cite{2026MNRAS.tmp..720F} and references therein) for which \cite{2020A&A...640A.106M} measured a mean age of 11.2 Gyr and metallicity [M/H] = --1.33 for the stars, which is consistent with the mean GC age of 9.1 Gyr and [M/H] = --1.44 within uncertainties. This is suggestive of either the GC system forming in the same event as the stellar body or the stellar body being comprised largely of disrupted GCs. Likely the answer is somewhat of both \citep{Danieli2022}.
   
   Here we find similarly for \glxA, that the GCs and the galaxy stars have the same ages and metallicities within uncertainties. Their ages are all approximately $\sim$9 Gyr and with a low metallicity of [M/H] $\sim$ --1. The mean galaxy colour of $g-r = 0.68$ is also consistent with the average for the stacked GCs. Again, this suggests that the GCs and stars formed at the same epoch from the same enriched gas, with very little subsequent enrichment within the galaxy. Over time, the stellar body would also have a  contribution from stars that had been disrupted from GCs \citep{ 2025MNRAS.536.1217F}. This process could alter the galaxy stellar populations to be similar to that of a metal-poor GC. For example, if the galaxy centre had achieved higher metallicities, then this process would effectively lower the galaxy's central metallicity, giving rise to the flat (\glxA) or even rising (\glxB) metallicity profiles seen in Fig. \ref{fig:stell_pop}.
   


\section{Conclusions}

Here we present KCWI spectra of two large-sized, low surface brightness galaxies in low density environments from the Dragonfly Ultrawide Survey. We derive  their global and radial stellar populations, along with their star formation histories. Both galaxies reveal overall old ages ($\sim$9 Gyr) and metal-poor ([M/H] $\sim$ --1) stars, which were quenched rapidly ($\sim$ 1 Gyr). They are thus both rare examples of a quiescent dwarf galaxy located in a low density environment. 
A further interesting feature is that both reveal flat-to-rising stellar population gradients. This behaviour is similar to that seen in ultra diffuse
galaxies but is in contrast with typical classical dwarf galaxies and predictions from simulations. 
For the galaxy (DFUWS 68) 
we find that its globular clusters 
share similar ages, metallicities,
and quenching timescales, to that of the host galaxy stars. This suggests that the star clusters and galaxy stars formed at the same epoch from the same enriched material and were quenched by the same mechanism. The brightest GC is located near to the galaxy centre and would probably be classified as a nuclear star cluster if it migrates to the centre. 
Neither cosmic web stripping nor internal feedback processes can fully explain the key properties presented in this work. 
More work is required to determine whether the large sizes, positive
radial gradients and short quenching timescales are all related.

\section*{Acknowledgements}
We thank the AGATE team, M. Collins and O. Newton for their help and useful comments.  We thank the referee for their comments. 
HC, DF and JB thank the ARC for financial support via DP250101673. AFM acknowledges support from RYC2021-031099-I and CEX2025-001609-S of , and together with DF and JG, PID2024-162088NB-I00 of MICIU/AEI. AJR was supported by National Science Foundation grant AST-2308390.

Software: {\tt pPXF, spectres, KCWI DRP, python, matplotlib, seaborn, astrophot, SExtractor}\\
Telescope: KCWI at Keck. Some of the data presented herein were obtained at Keck Observatory, which is a private 501(c)3 non-profit organization operated as a scientific partnership among the California Institute of Technology, the University of California, and the National Aeronautics and Space Administration. The Observatory was made possible by the generous financial support of the W. M. Keck Foundation. The authors wish to recognize and acknowledge the very significant cultural role and reverence that the summit of Maunakea has always had within the Native Hawaiian community. We are most fortunate to have the opportunity to conduct observations from this mountain.

\section*{Data Availability}

The data used in this project are publicly available via the Keck Observatory Archive. 
 



\bibliographystyle{mnras}
\bibliography{bibliography} 




\appendix

\section{Legacy imaging}
\label{app:light-weighted}

\begin{figure*}
    \centering
    \includegraphics[width=0.95\linewidth]{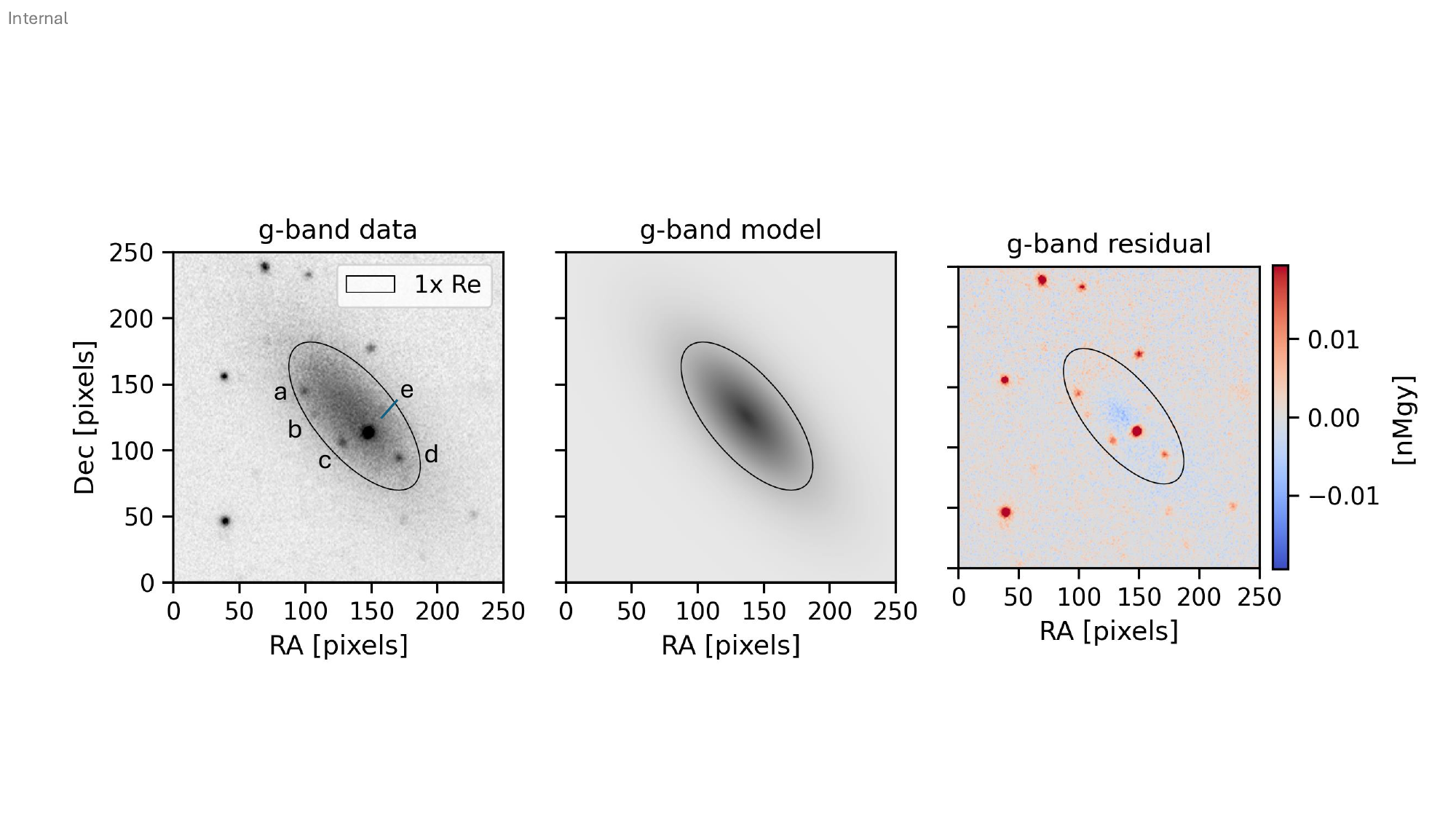}
    \caption{Legacy imaging of \glxA\ in the g-band with one effective radius ellipse shown. The Legacy pixel scale is 0.26 arcsec. {\it Left} Original Legacy image, with GC IDs indicated. {\it Middle} Galaxy Sersic model (which excludes point sources). {\it Right} Residual image of original minus model revealing the point sources. This residual image is used to determine the point source magnitudes.
    }
    \label{fig:legacy}
\end{figure*}

\section{Light-weighted results}
\begin{figure}
    \centering
    \includegraphics[width=\linewidth]{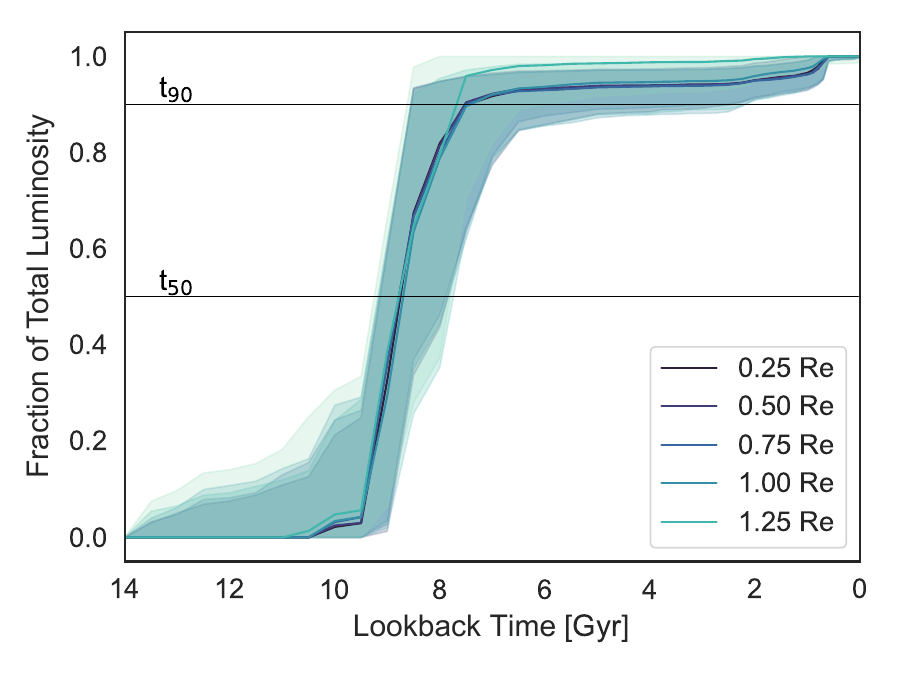}
    \includegraphics[width=\linewidth]{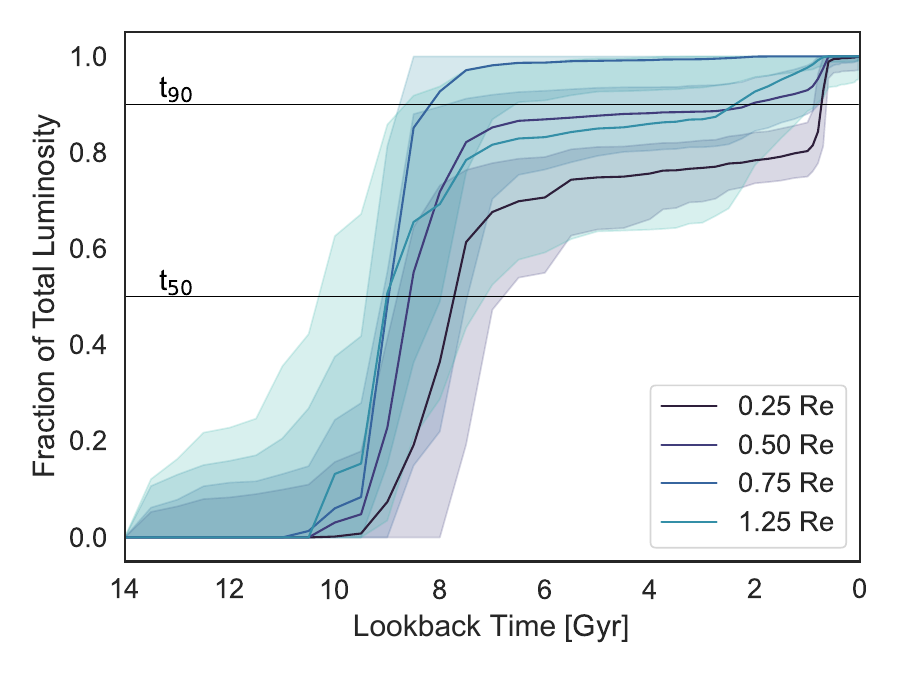}
    \caption{
Light-weighted star formation history in various radial bins  \glxA\ {\it (upper)} and \glxB\ {\it (lower)}.
    Shaded regions show the uncertainty. Horizontal lines represent the time to acquire 50\% and 90\% of the total stellar mass of the galaxy. 
    }
    \label{fig:sfh_lw}
\end{figure}

\begin{figure}
    \centering
    \includegraphics[width=\linewidth]{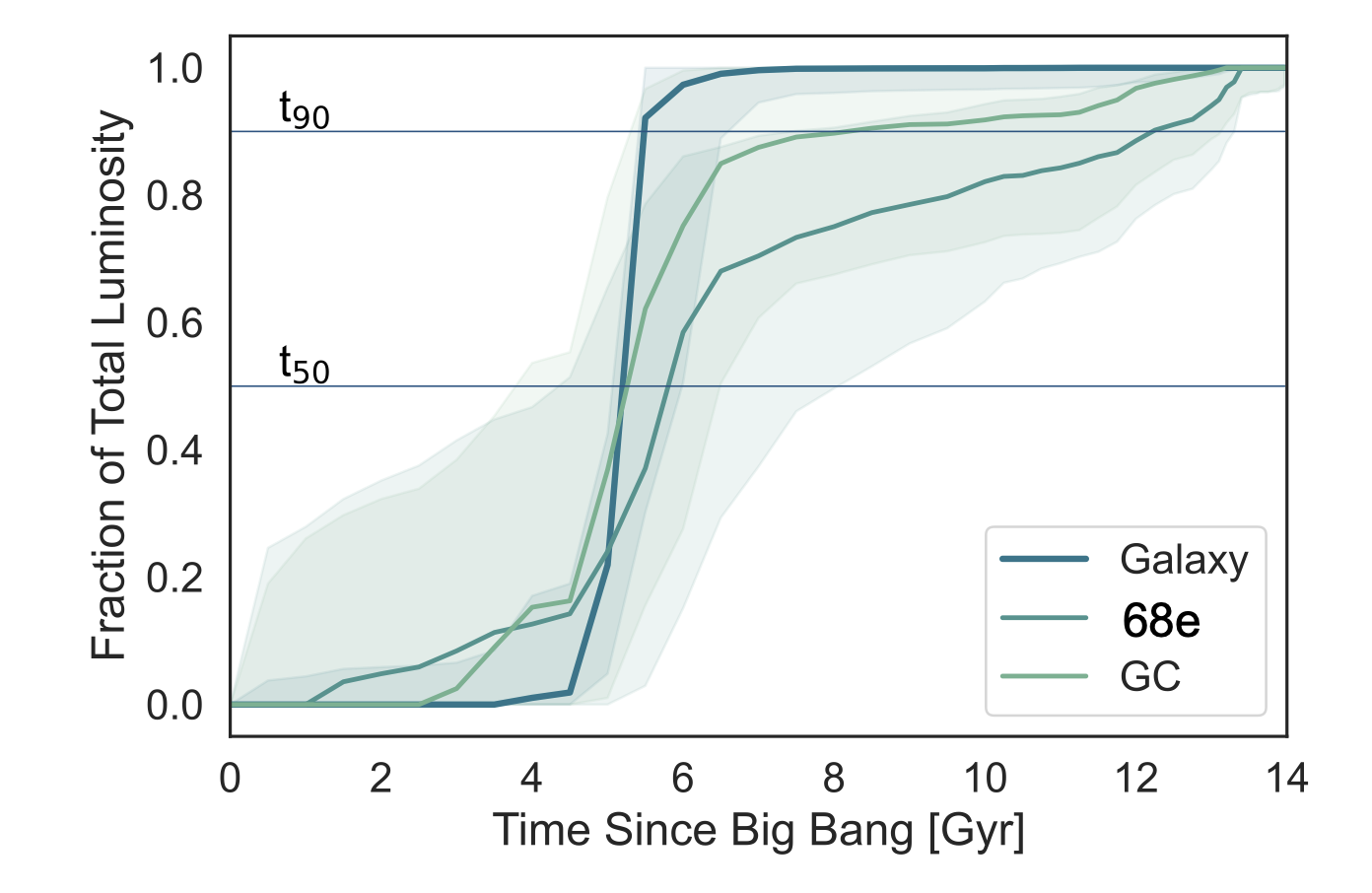}
    \caption{
Light-weighted star formation history for the brightest GC and possible nuclear star cluster (86e), the combined GCs and the host galaxy \glxA\ .
    Shaded regions show the uncertainty. Horizontal lines represent the time to acquire 50\% and 90\% of the total stellar mass of the galaxy. 
 }   
    \label{fig:placeholder}
\end{figure}

\section{Radial Bins}

\begin{figure}
    \centering
    \includegraphics[width=\linewidth]{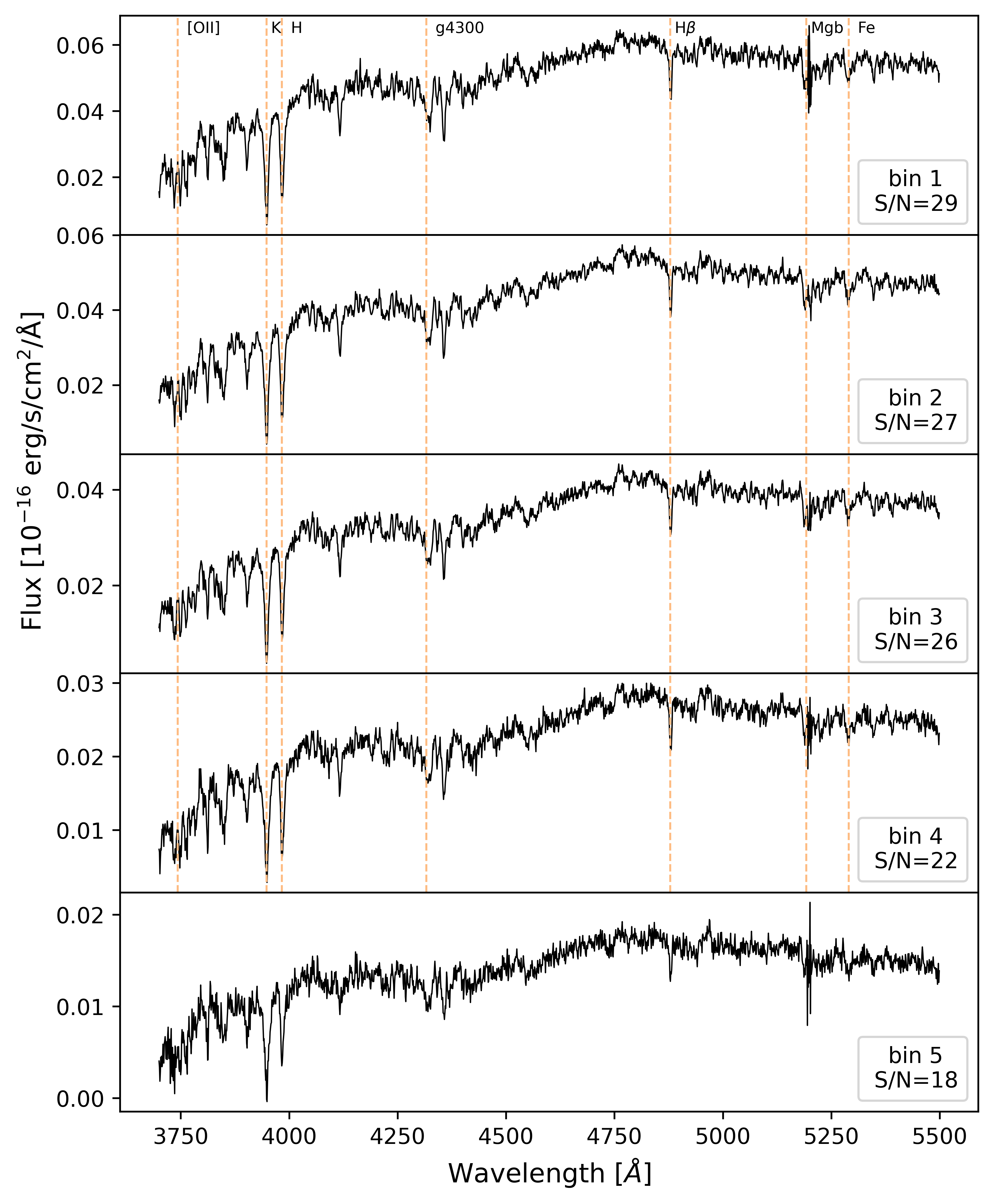}
    \caption{Spectra in radial bins 1--5 to a radius of 1.25~R$_e$ for \glxA. In each panel the main spectral lines are indicated, along with the S/N   in each radial bin. 
 }   
    \label{fig:bins68}
\end{figure}

\begin{figure}
    \centering
    \includegraphics[width=\linewidth]{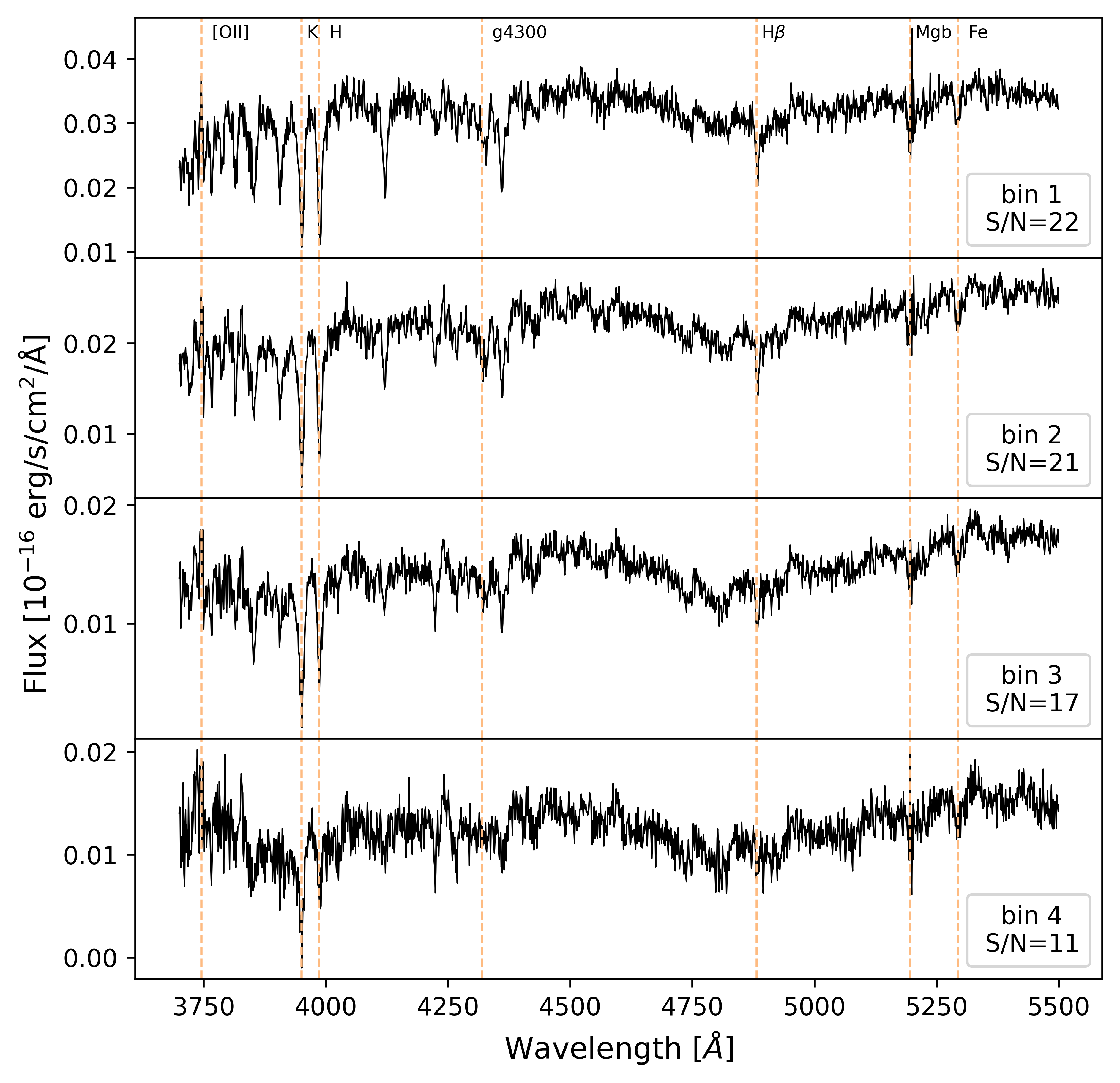}
    \caption{Spectra in radial bins 1--4 f
    to a radius of 1.25~R$_e$ 
    for \glxB. In each panel the main spectral lines are indicated, along with the S/N  in each radial bin. 
 }   
    \label{fig:bins258}
\end{figure}





\bsp	
\label{lastpage}
\end{document}